\documentclass[prl,preprint,superscriptaddress,showpacs]{revtex4-1}
\usepackage{amsmath}
\usepackage{amssymb}
\usepackage{geometry}                 
\usepackage{graphicx}
\usepackage[utf8]{inputenc}
\usepackage[english]{babel}
\usepackage{dsfont}

\begin{document}

\title{Knotted Strange Attractors and Matrix Lorenz Systems}
\author{Julien Tranchida}
\email{julien.tranchida@cea.fr}
\affiliation{CEA DAM/Le Ripault, BP 16, F-37260, Monts, FRANCE}
\affiliation{CNRS-Laboratoire de Math\'ematiques et Physique Th\'eorique (UMR 7350), F\'ed\'eration de Recherche "Denis Poisson" (FR2964), D\'epartement de Physique, Universit\'e de Tours, Parc de Grandmont, F-37200, Tours, FRANCE}
\author{Pascal Thibaudeau}
\email{pascal.thibaudeau@cea.fr}
\affiliation{CEA DAM/Le Ripault, BP 16, F-37260, Monts, FRANCE}
\author{Stam Nicolis}
\email{stam.nicolis@lmpt.univ-tours.fr}
\affiliation{CNRS-Laboratoire de Math\'ematiques et Physique Th\'eorique (UMR 7350), F\'ed\'eration de Recherche "Denis Poisson" (FR2964), D\'epartement de Physique, Universit\'e de Tours, Parc de Grandmont, F-37200, Tours, FRANCE}

\date{\today}                                           

\begin{abstract}
A generalization of the Lorenz equations is proposed where the variables take values in a Lie algebra. The finite dimensionality of the representation  encodes the quantum fluctuations, while the non-linear nature of the equations can describe  chaotic fluctuations. We identify a criterion, for the appearance  of such non-linear terms. This depends on whether an invariant, symmetric tensor of the algebra can vanish or not. This proposal is studied in detail  for the fundamental representation of $\mathfrak{u}(2)$. We find a knotted structure for  the attractor, a bimodal distribution for the largest Lyapunov exponent and that the dynamics takes place within the Cartan subalgebra, that does not contain only the identity matrix, thereby can describe the quantum fluctuations.  
\end{abstract}


\maketitle
Recent advances in magnetic materials and techniques allow manipulation of spin moments at nanoscale resolution. Thus, chaotic fluctuations become 
significant and their control have been the subject of both experimental and theoretical extensive studies \cite{Wigen:1994hc}. One of the first direct observation of period doubling and chaos was spin-wave instabilities in yttrium iron garnet (YIG) and has been documented more than thirty years ago \cite{Gibson:1984fu}. Using the ferromagnetic resonance technique, several routes to chaos have been found and explored including periodic-doubling cascades, quasi periodic and intermittent dynamics which exhibit complex magnetic behaviors \cite{Fernandez-Alvarez:2000ij,Mayergoyz:2009kl}. Recently, the phase diagram of a chaotic magnetic nanoparticle has been presented \cite{Bragard:2011nx}, which was obtained by monitoring the classical dynamics of its magnetization, modeled by the Landau-Lifshitz-Gilbert (LLG) equation.

What has received much less attention is the contribution of {\em quantum} fluctuations, that become significant at nanoscale resolution and are 
crucial for controlling qubit devices~\cite{boissoneault}. These might affect non-linear effects in new ways.

The challenge, therefore, is to describe the interplay between the two sources of fluctuations in a way that can lead to a deeper understanding of their effects and predict new features. To this end a model is proposed that displays both and allows us to distinguish them in a particularly clean fashion.

To take into account the quantum fluctuations, non-commuting variables are used and taken to be finite-dimensional matrices. To describe chaotic 
fluctuations we will avail ourselves of the universality classes of dynamical systems, described by ordinary differential equations.

\begin{figure}[thp]
\resizebox{0.9\columnwidth}{!}{\includegraphics{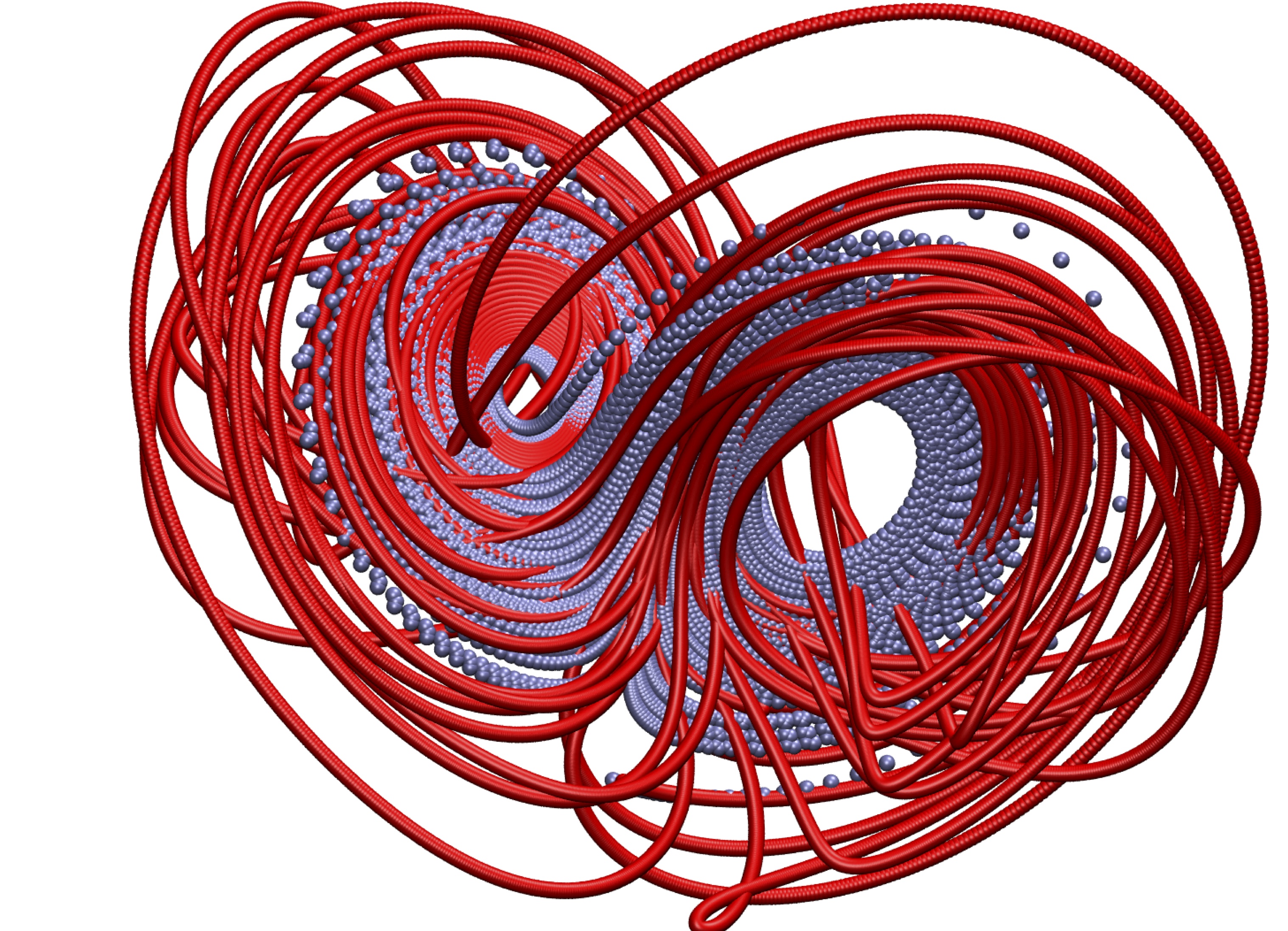}}
\caption{(Color online) Plot of the parametric solution of the time evolution of $\mathrm{Tr}(X), \mathrm{Tr}(Y)\, \mathrm{and}\, \mathrm{Tr}(Z)$ 
with $\sigma=10$, $r=28$, $b=8/3$. Classical Lorenz system in gray and its matrix version in red. \label{Fig1}}
\end{figure}

The prototypical system that displays the full repertoire of behaviors from regular to chaotic, along many routes, is the Lorenz 
system~\cite{Lorenz:1963vn}
\begin{equation}
\begin{array}{ccl}
\dot{x}&=&\sigma(y-x)\\
\dot{y}&=&x(r-z)-y\\
\dot{z}&=&xy-bz
\end{array}
\label{CLS}
\end{equation}
whose typical solution (in the chaotic phase) is displayed in Fig.~\ref{Fig1}. 
Here $\dot{x}\equiv dx/dt$ and similarly for the other variables. While these equations were deduced to describe classical fluid dynamics and the 
original parameters reflect this fact: $\sigma$ is the Prandtl number, $r$ the Rayleigh number and $b$ is the aspect ratio of the ``cell'' (real or 
virtual), their scope is, in fact, much broader, as was realized from the work in the 70's~\cite{Feigenbaum:1978fk}. These are, still, classical 
equations and the variables $(x,y,z)$ are commuting quantities. 

In order to describe quantum fluids, a generalization of these equations to the case where the variables become operators, $(X,Y,Z)$ is the natural 
way~\cite{Floratos:2012ys}. Our starting point will be the model discussed in ref.~\cite{Axenides:2010zr}, where the operators are described by square matrices, that are 
expanded in the generators of a given Lie algebra. As is usual in quantum mechanics, a prescription for the expressions that involve products of 
non-commuting variables is mandatory. The Weyl ordering~\cite{Balazs:1984uq} is adopted, so that  the product $XY$ of two operators, $X$ and $Y$  is replaced by its symmetric expression $XY\to \frac{1}{2}\left(XY + YX\right)$. The  $X,Y,Z$ are now expanded in the generators, $T^a$, of a Lie algebra, $\mathcal{G}$, where $a=1,2,\ldots,\mathrm{dim}_{\mathrm{Ad(\mathcal{G})}}$, thus $X\equiv x^a T^a, Y\equiv y^a T^a ~$and$~ Z\equiv z^a T^a$. 
A Lie algebra is defined by its structure constants, $f^{abc}$, that enter in the commutation relations, $[T^a,T^b]=\mathrm{i}f^{abc}T^c$ and the 
fact that this algebra is compact implies that $\mathrm{Tr}( T^a T^b)= \kappa\delta^{ab}$ with $\kappa\neq 0$ and $\mathrm{Tr}$ stands for the trace on the algebra. Using such requirements, the quantum counterparts to eqs.~(\ref{CLS}) take the following, intriguing, form
\begin{equation}
\begin{array}{ccl}
\dot{x}^a&=&\displaystyle{\sigma (y^a - x^a)}\\
\dot{y}^a&=&\displaystyle{-y^a + r x^a -d^{abc}x^b z^c}\\
\dot{z}^a&=&\displaystyle{-bz^a + d^{abc}x^b y^c}
\end{array}
\label{MLS}
\end{equation}
These equations, herald the appearance of  the invariant symmetric tensor
\begin{equation}
\label{anomaly_tensor}
d^{abc}\equiv\frac{1}{2\kappa}\mathrm{Tr}\left[\left\{T^a,T^b\right\}T^c\right]
\end{equation}
of the Lie algebra. This tensor appears in gauge theories, since the gauge fields belong to the adjoint representation of the group and plays an 
important role in the classification of gauge anomalies~\cite{Georgi:1972fk}. Groups, for which this tensor vanishes identically are 
called ``anomaly--safe'' and in the present context, this means that eqs.~(\ref{MLS}) are {\em linear}, thus do not give rise to chaotic fluctuations. 
Only quantum fluctuations can appear. Groups, for which this tensor does not identically vanish, on the other hand, lead to non--linear equations, 
thus can describe both chaotic {\em and} quantum fluctuations. 


For the case of abelian Lie groups, the structure constants vanish. In that case, eqs~(\ref{MLS}) is equivalent to eqs~(\ref{CLS}) up to a rescaling 
of all the variables proportional to the single non-vanishing element of $d$. This means that they share the same route to chaos, i.e. belong to the 
same universality class. 

In order of complexity, the case of ``anomaly--safe'' groups comes next. For them $d^{abc}=0$ identically, so chaotic fluctuations are absent and 
only quantum fluctuations remain,  which can be consistently identified. The simplest case is that of the $\mathfrak{su}(2)$ algebra, that is relevant for the 
description of the magnetization operator, where the variables $X, Y$ and $Z$ are considered to be the components of the magnetization of a quantum 
system. In ref~\cite{Roupas:2012fk} it is noted that the Lorenz equations do, in fact, describe the evolution of the magnetization of a magnetic 
top, subject to a magnetic field, that controls the dissipation. The matrix equations may provide a consistent quantization of this case. 

When the $d$ tensor no longer vanishes, chaotic fluctuations and quantum effects mix in a non-trivial way. The simplest case is the 
$\mathfrak{u}(2)=(\mathfrak{u}(1)\times \mathfrak{su}(2))/\mathbb{Z}_2$ algebra, where all the components of $d^{abc}$ are zero, except $d^{0cc}=1/2$ for $c\in\{0,1,2,3\}$ with the 
corresponding circular permutations. This can be made particularly clear by writing the corresponding Lorenz system of equations as follows:

\begin{equation}
\label{SU2U1}
\begin{array}{l}
\mathfrak{u}(1)
\left\{
\begin{array}{l}
\dot{x}^0=\sigma(y^0-x^0)\\
\dot{y}^0=-y^0+r x^0 -x^0z^0-2 x^b z^b\\
\dot{z}^0=-b z^0 + x^0 y^0 + 2x^b y^b
\end{array}
\right.
\\
\\
\mathfrak{su}(2)
\left\{
\begin{array}{l}
\dot{x}^a=\sigma(y^a-x^a)\\
\dot{y}^a = -y^a+r x^a-x^0 z^a-x^a z^0\\
\dot{z}^a = - bz^a + x^a y^0 + x^0 y^a
\end{array}
\right.
\end{array}
\end{equation}

The first set of the three equations highlights the fact that $\mathfrak{u}(1)$ is responsible for the chaotic fluctuations and the second set shows that 
the $\mathfrak{su}(2)$ components satisfy linear equations, with the $\mathfrak{u}(1)$ variables acting as sources. 

The two sets of equations were integrated using an eighth-order Runge-Kutta scheme with a fixed time stepping scheme. The group invariants, $\mathrm{Tr}(X)$, $\mathrm{Tr}(Y)$ and $\mathrm{Tr}(Z)$, plotted in Fig.~\ref{Fig1}, define a subspace where a structure similar to the Lorenz ``butterfly'' appears, ``decorated'' by the $\mathfrak{su}(2)$ terms that make it ``knotted''. This last property is, in fact, expected \cite{Ghrist:1998uq}, given that great circles on the $\mathfrak{su}(2)$ manifold, the 3--sphere, do have non-zero linking number, the well known Hopf invariant~\cite{Niemi:2014kx}. 

\begin{figure}[htb]
\resizebox{0.9\columnwidth}{!}{\includegraphics{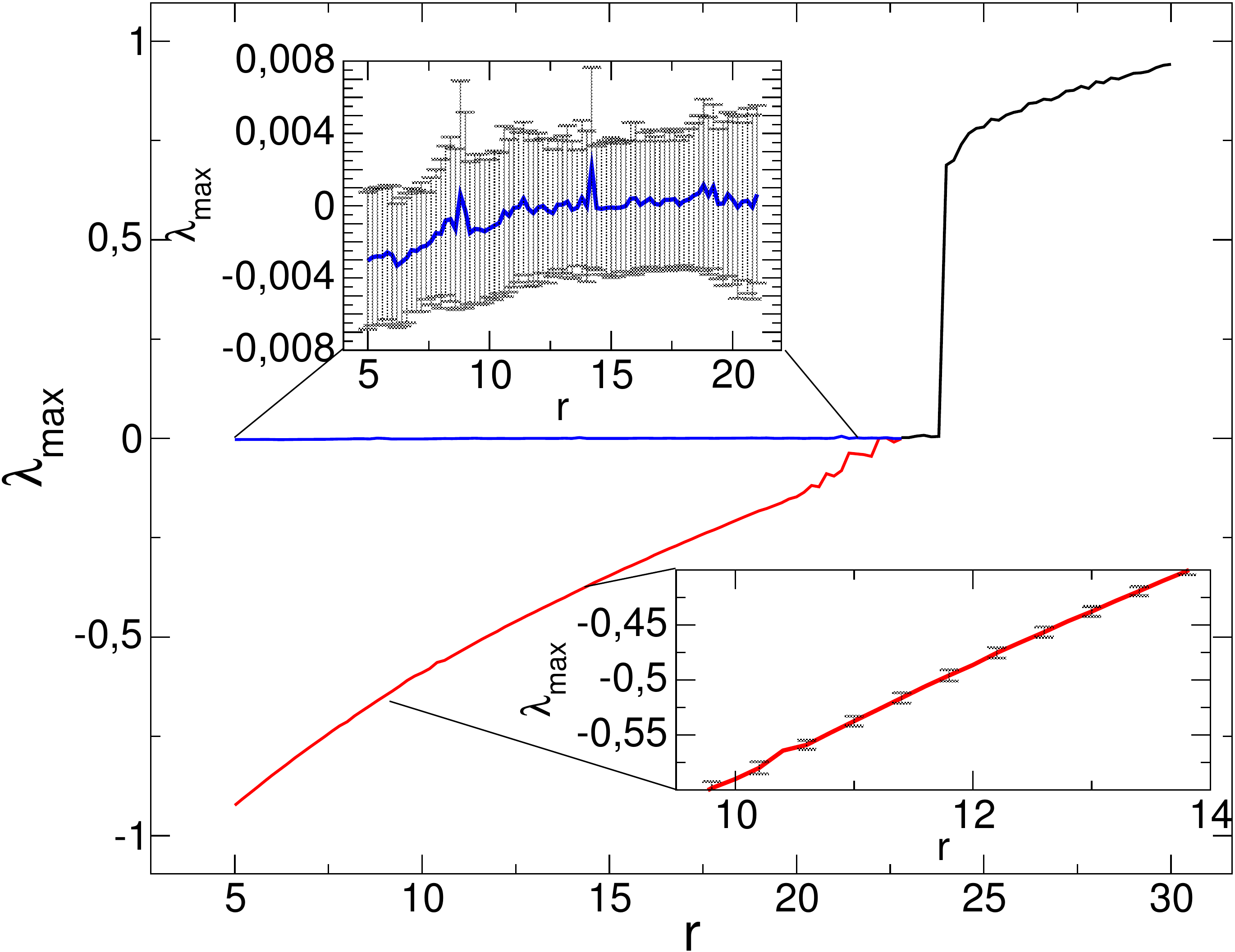}}
\caption{(Color online) Phase diagram of the largest Lyapunov spectra in the $\mathfrak{u}(2)$ matrix Lorenz system. Details of selected parts of the diagram are inserted to exhibit the 1$\sigma$ error bars. \label{Fig2}} 
\end{figure}

To monitor the chaotic fluctuations, the full Lyapunov spectrum is computed, using the method of Christiansen and Rugh~\cite{CR} and comprises of 3$\times$dim$_{\mathrm{Ad}(\mathcal{G})}=12$ exponents. A useful proxy for their behavior is given by the distribution of the maximal Lyapunov exponent, taken over random  initial conditions, for fixed values of the parameters. $\lambda_{max}$ is plotted in Fig.~\ref{Fig2} as a function of $r$, for fixed $b$ and $\sigma$. Each point is an average over two hundred runs with random initial conditions. Particular care must be exercised in avoiding the, rugged, stable manifold, not only, of the origin~\cite{Sparrow:1982uq}, but, also, of the other, linearly stable structures, zeros of the right hand side of eqs.~(\ref{SU2U1}). 

A striking feature of the largest Lyapunov exponent, as function of $r$, is its bimodal distribution, along the two group factors, 
$\mathfrak{u}(1)$ and $\mathfrak{su}(2)$, in the non-chaotic phase, $r<r_\mathrm{crit}$, whereas, in the chaotic phase, $r>r_\mathrm{crit}$, the distribution 
becomes unimodal, giving the same values for each group factor. While a mathematical proof for this result is not available, we stress that it provides a consistency check for the reliability of our numerical analysis, since the $\mathfrak{su}(2)$ factor cannot give rise to chaos by itself. 
The transition to chaos, at $r=r_\mathrm{crit}$, appears at the same value as for the classical Lorenz system, only much more abrupt in the matrix case. 

\begin{figure}[htb]
\resizebox{0.9\columnwidth}{!}{\includegraphics{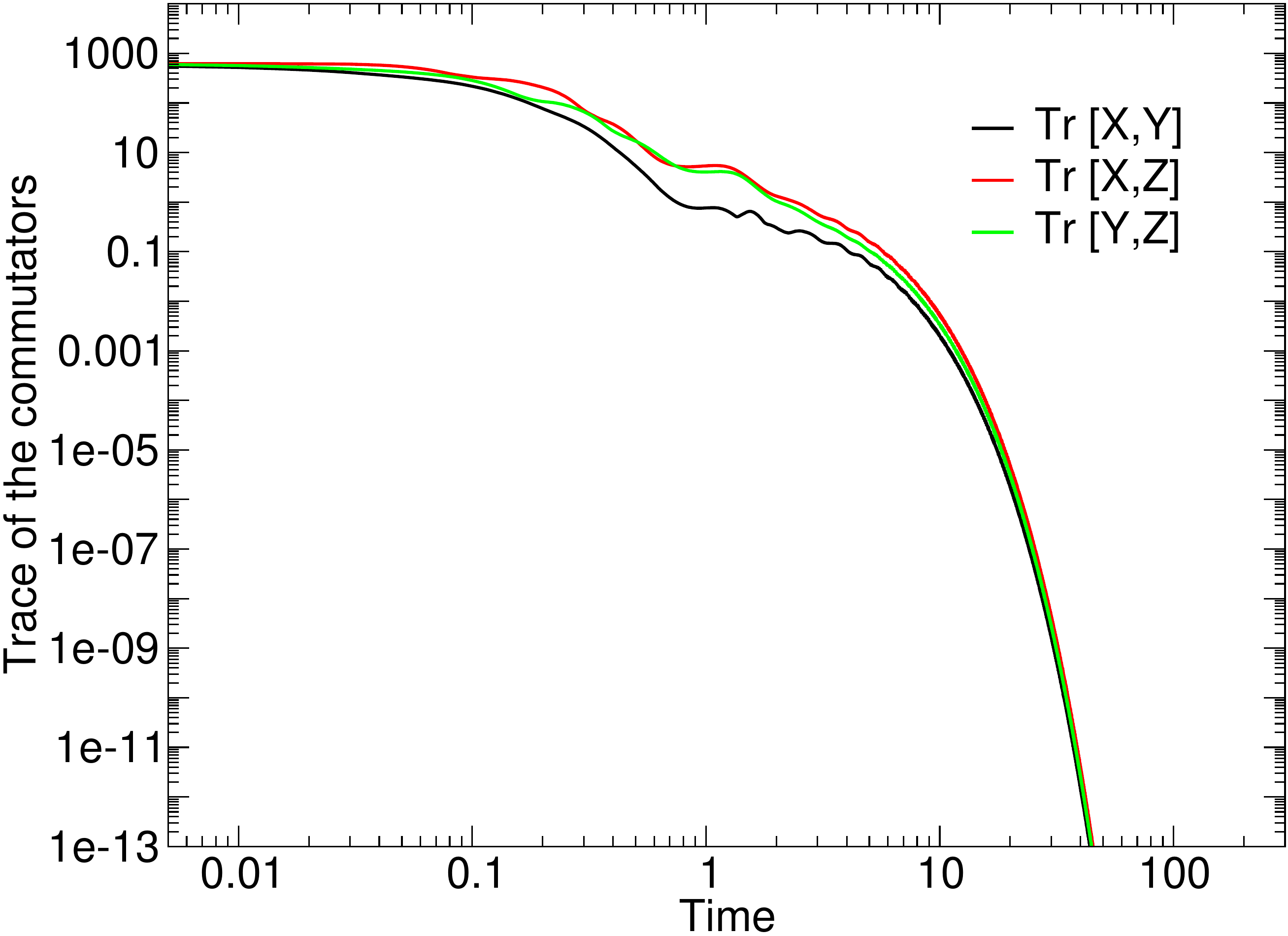}}
\caption{(Color online.) Plot of the average of the three commutators ${\sf{Tr}}([X,Y]),~{\sf{Tr}}([Y,Z]),~{\sf{Tr}}([X,Z])$ taken over random initial conditions for $r=15$. \label{LorMatplot}}
\end{figure}

Another way to characterize the quantum fluctuations is by studying the time evolution of the three commutators, ${\mathrm{Tr}}([X,Y])$, ${\mathrm{Tr}}([Y,Z])$ and ${\mathrm{Tr}}([X,Z])$. If all vanish, this means that all three matrices belong to the Cartan subalgebra. Preliminary results show that this does, in fact, occur~\cite{Axenides:2010zr} and our numerical results seem to confirm it.  What is noteworthy in Fig.~\ref{LorMatplot} is that the time evolution of the average of the three commutators taken over random initial conditions, collapses along the same curve.
We shall now eliminate the possibility that the Cartan subalgebra so obtained is proportional only to the identity. This is achieved by monitoring the invariants in the $\mathfrak{su}(2)$ subspace, namely $\mathrm{Tr}X^2-{(\mathrm{Tr}X)}^2={x_1}^2+{x_2}^2+{x_3}^2$ and so on. Performing runs for different values of $r$, at fixed $b$ and $\sigma$, we find that, in the non-chaotic phase, $r<r_\mathrm{crit}$, these quantities seem to converge to a single point, not the origin, whereas in the chaotic phase, $r>r_\mathrm{crit}$, they appear to describe a fuzzy region, of finite volume in phase space. That it does not collapse to the origin indicates the persistence of the quantum fluctuations in both phases. 

A natural generalization of our results is towards a quantum Nambu description of spin systems, where it has been demonstrated that the Nambu mechanics leads to novel identities for extended Lorenz system with dissipation that are not obvious in an Hamiltonian approach \cite{Blender:2013kx}. Moreover this generalization may also provide insights into  the origin of  dissipation in such magnetic systems that has become a subject of topical research~\cite{garanin1997fokker} and lead to new relations that are hard to guess from the Hamiltonian viewpoint. 

\begin{acknowledgments}
JT acknowledges financial support through  a joint doctoral fellowship ``R\'egion Centre-CEA''.   
\end{acknowledgments}

\bibliographystyle{apsrev4-1}
\bibliography{lorenz}

\end{document}